\documentclass{ws-ijmpe}

\begin{document}

\newcommand{\Slash}[1]{\ooalign%
{\hfil\rotatebox{30}{\underline{\hspace*{0.6cm}}}\hfil\crcr$#1$}}

\markboth{M.~Harada and C.~Sasaki}
{Thermal Dilepton Production from Dropping $\rho$ 
in the Vector Manifestation}

\catchline{}{}{}{}{}

\title{Thermal Dilepton Production from Dropping $\rho$\\ 
in the Vector Manifestation
}

\author{\footnotesize Masayasu Harada}

\address{Department of Physics, Nagoya University,
Nagoya, 464-8602, Japan\\
harada@hken.phys.nagoya-u.ac.jp}

\author{Chihiro Sasaki}

\address{ GSI, D-64291 Darmstadt, Germany\\
c.sasaki@gsi.de}

\maketitle

\begin{history}
\received{(received date)}
\revised{(revised date)}
\end{history}

\begin{abstract}
We study the pion electromagnetic form factor and the dilepton
production rate in hot matter based on
the vector manifestation (VM) of chiral symmetry 
in which the massless vector meson becomes the chiral partner 
of the pion, giving a theoretical framework of the dropping 
$\rho$ \^^ a la Brown-Rho scaling.
The VM predicts a strong violation of the vector dominance (VD) 
near the phase transition point associated with the dropping $\rho$.
We present the effect of the VD violation to the dilepton 
production rate and make a comparison to the one predicted 
by assuming the VD together with the dropping $\rho$.
\end{abstract}

\section{Introduction}
\label{sec:int}

An enhancement of dielectron mass spectra below
the $\rho / \omega$ resonance was first observed at CERN 
SPS~\cite{ceres} and it is considered as an indication of 
the medium modification of the vector mesons.
Still the vector meson mass in matter remains an open issue~
\cite{KEK-PS,trnka,SB:STAR,NA60}.
Although several scenarios
like collisional broadening due to interactions with the
surrounding hot and dense medium~\cite{RW} 
and dropping $\rho$ meson mass associated with chiral symmetry 
restoration~\cite{BR-scaling,HY:VM} have been discussed,
there are conceivable ambiguities which have not been
considered~\cite{Brown:2005ka-kb,HR,SG} and
no conclusive distinction between them has been done.

The vector manifestation (VM)~\cite{HY:VM,HY:PRep} was proposed
as a novel pattern of the Wigner realization of chiral symmetry
with a large number of massless quark flavors by using
the hidden local symmetry (HLS) theory,
in which the vector meson becomes massless at the restoration
point and belongs to the same chiral multiplet as the pion,
i.e., {\it the massless vector meson is the chiral partner 
of the pion}.
The studies of the VM in hot/dense matter have been carried out~
\cite{HS:VM,HKR:VM,VM:T} and the VM was also
applied to construct an effective Lagrangian for the heavy-light 
mesons which can well describe the recent experimental observation 
on the $D(0^+,1^+)$ mesons~\cite{HRS:CD}.

It has been shown that the vector dominance (VD) 
of the electromagnetic form factor of the pion~\cite{Sakurai}
is accidentally satisfied in $N_f=3$ QCD 
at zero temperature and zero density, and that it is strongly 
violated in large $N_f$ QCD when the VM occurs~\cite{HY:VD}.
The VD is characterized by the direct $\gamma\pi\pi$ being zero.
In hot/dense matter, the $\gamma\pi\pi$ coupling is modified
by medium effects and approaches 1/2 with increasing $T/\mu_q$ 
toward the critical point, which is due to the intrinsic
$T/\mu_q$ effects associated with the chiral symmetry restoration
~\cite{HKR:VM,VM:T}.
This implies that {\it the VD is strongly violated near the critical 
point, maximally by 50 \%}.
It strongly affects the understanding of experiments
on the dilepton productions based on the dropping $\rho$~
\cite{Brown:2005ka-kb},
and recently a quantitative study has been done~\cite{HS:dilepton}.
In this contribution,
we focus on the electromagnetic form factor of pion and 
the dilepton production rate at finite temperature
from the dropping $\rho$ based on the VM
with paying a special attention to the effect of the violation 
of the VD following Ref.~\refcite{HS:dilepton}


\setcounter{equation}{0}
\section{Form factor and dilepton spectra}
\label{sec:FF}

In Ref.~\refcite{HY:PRep},
the bare parameters of the HLS Lagrangian were determined
by performing the Wilsonian matching, in which
the axial-vector and vector current correlators derived from the
HLS with those by the operator product expansion in
QCD are matched at a matching scale $\Lambda \sim 1$ GeV.
This procedure in hot matter provides the following proportionality 
between the $\rho$-meson mass and the chiral condensate
near $T_c$~\cite{HS:VM}:
\begin{eqnarray}
m_\rho(T) \propto \langle \bar{q}q \rangle_{T}\,,
\label{match-input}
\end{eqnarray}
This implies that $m_\rho$ is thermally evolved
following the temperature dependence of the quark condensate,
which is nothing but the intrinsic temperature effect.

It should be stressed that
Eq.~(\ref{match-input}) holds
{\it only in the vicinity of $T_c$} and
is not valid any more far away from
$T_c$ where ordinary hadronic corrections are dominant.
For expressing a temperature above which the intrinsic
effect becomes important,
we shall introduce a temperature $T_f$, 
so-called flash temperature~\cite{BLR:flash,BLR}.
The VM and therefore the dropping $\rho$ mass become 
transparent for $T>T_f$.
On the other hand, we expect that
the intrinsic effects are negligible in the low-temperature
region below $T_f$:
Only hadronic thermal corrections are considered for $T < T_f$.
Here we would like to remark that
the Brown-Rho scaling deals with the quantity directly locked to the 
quark condensate and hence {\it the scaling masses are achieved 
exclusively by the intrinsic effect} in the present framework.

A lepton pair is emitted from the hot/dense matter
through a decaying virtual photon.
The differential production rate in the medium for fixed 
temperature $T$
is expressed in terms of the imaginary part of the photon 
self-energy $\mbox{Im}\Pi$ as
\begin{equation}
\frac{dN}{d^4q}(q_0,\vec{q};T)
=\frac{\alpha^2}{\pi^3 M^2}\frac{1}{e^{q_0/T}-1}
\mbox{Im}\Pi (q_0,\vec{q};T)\,,
\label{rate}
\end{equation}
where $\alpha = e^2/4\pi$ is the electromagnetic coupling constant,
$M$ is the invariant mass of the produced dilepton and 
$q_\mu=(q_0,\vec{q})$ denotes the momentum of the virtual photon.
We will focus on an energy region around the $\rho$ meson mass
scale in this analysis.
In this energy region it is natural to expect that
the photon self-energy is dominated by the two-pion process
and its imaginary part is related to the pion electromagnetic 
form factor ${\cal F}(s;T)$ through 
\begin{equation}
\mbox{Im}\Pi(s;T)
= \frac{1}{6\pi\sqrt{s}}
\left( \frac{s - 4m_\pi^2}{4} \right)^{3/2}
\left| {\cal F}(s;T) \right|^2\,,
\label{Im Pi}
\end{equation}
where $s$ is the square of the invariant mass
and $m_\pi$ is the pion mass.

\begin{figure}
\begin{center}
\includegraphics[width = 5cm]{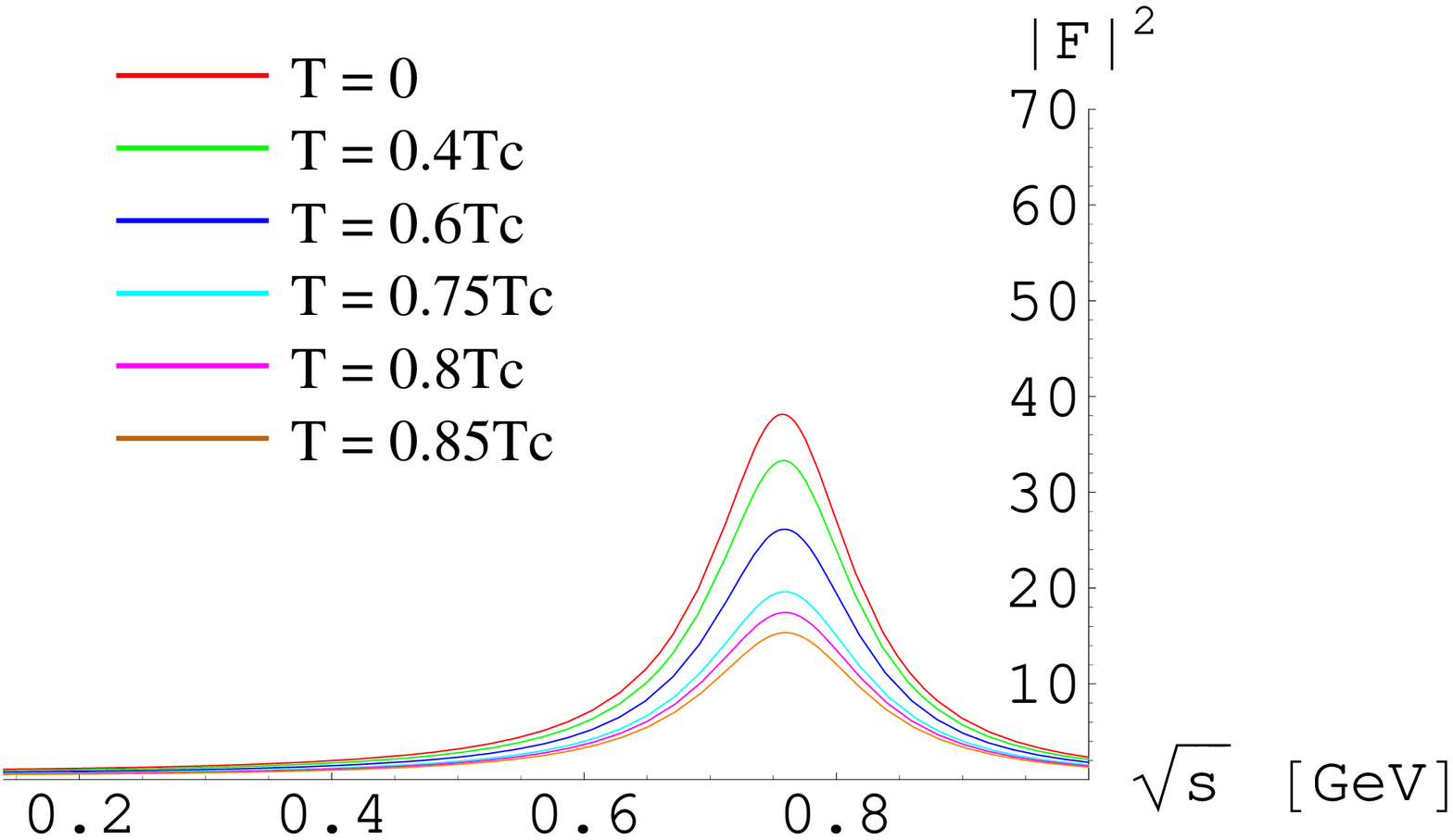}
\hspace*{0.5cm}
\includegraphics[width = 5cm]{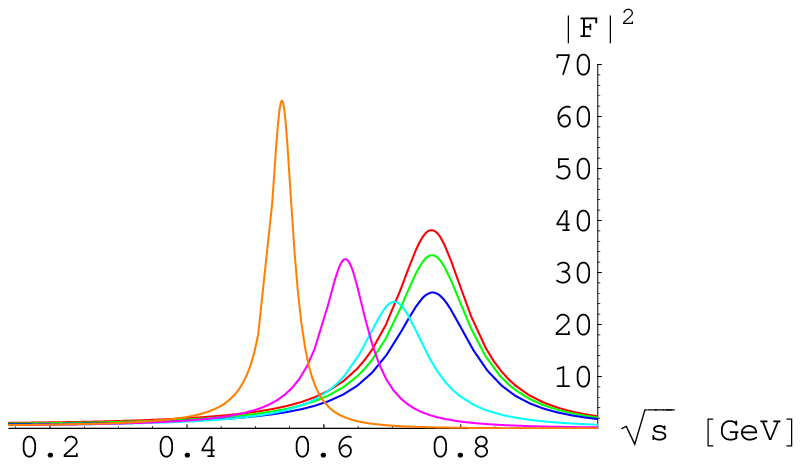}
\\
(a)
\hspace*{5cm}
(b)
\end{center}
\caption{
Electromagnetic form factor of the pion as a function of
the invariant mass $\sqrt{s}$ for several temperatures.
The curves in the left panel (a) include only the hadronic
temperature effects and those in the right panel (b) include
both intrinsic and hadronic
temperature effects.
}
\label{fig:form}
\end{figure}

Figure~\ref{fig:form} shows the pion electromagnetic 
form factor ${\cal F}$ for several temperatures.
In Fig.~\ref{fig:form} (a) 
there is no remarkable shift of the $\rho$ meson mass
but the width becomes broader with increasing temperature, 
which is consistent with the previous study in Ref.~\refcite{SK}.
In Fig.~\ref{fig:form} (b) the intrinsic temperature effect are also
included into all the parameters in the form factor.
At the temperature below $T_f$, 
the hadronic effect dominates the form factor,
so that the curves for $T = 0$, $0.4T_c$ and $0.6T_c$
agree with the corresponding ones in Fig.~\ref{fig:form}(a).
At $T = T_f$ the intrinsic effect starts to contribute
and thus in the temperature region above $T_f$ 
the peak position of the form factor moves as
$m_\rho(T) \rightarrow 0$ with increasing temperature toward $T_c$.
Associated with this dropping $\rho$ mass,
the width becomes narrow and 
the value of the form factor at the peak
grows up~\cite{HS:VM}.

\begin{figure}
\begin{center}
\includegraphics[width = 5cm]{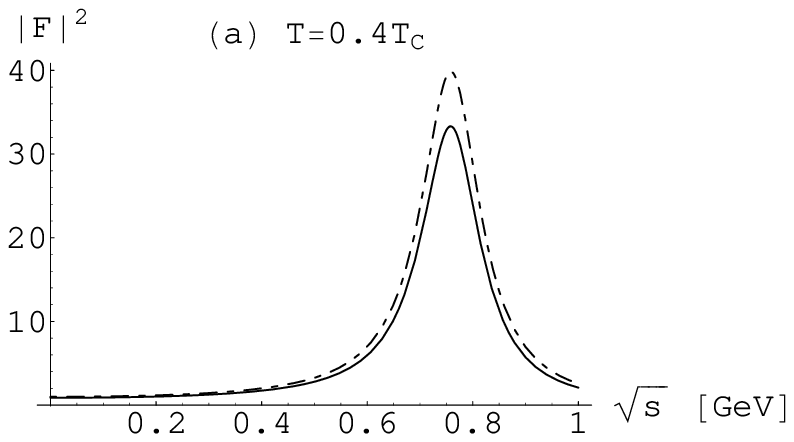}
\hspace*{0.5cm}
\includegraphics[width = 5cm]{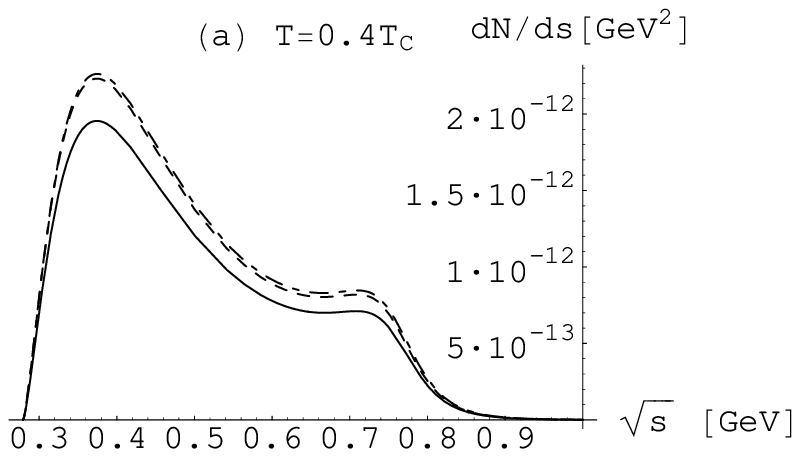}
\\
\includegraphics[width = 5cm]{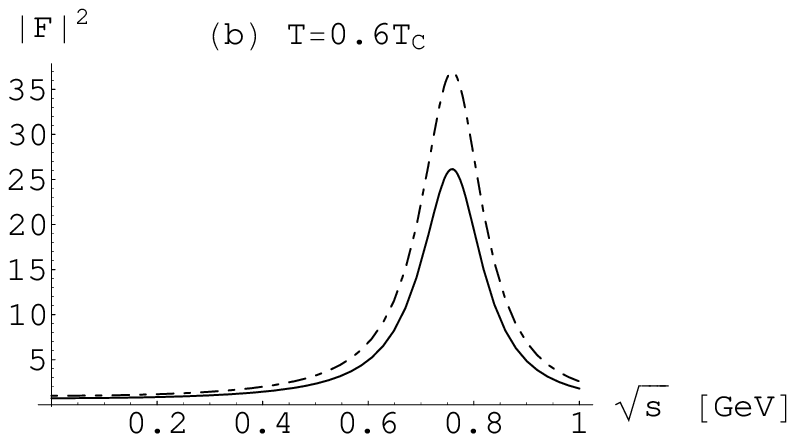}
\hspace*{0.5cm}
\includegraphics[width = 5cm]{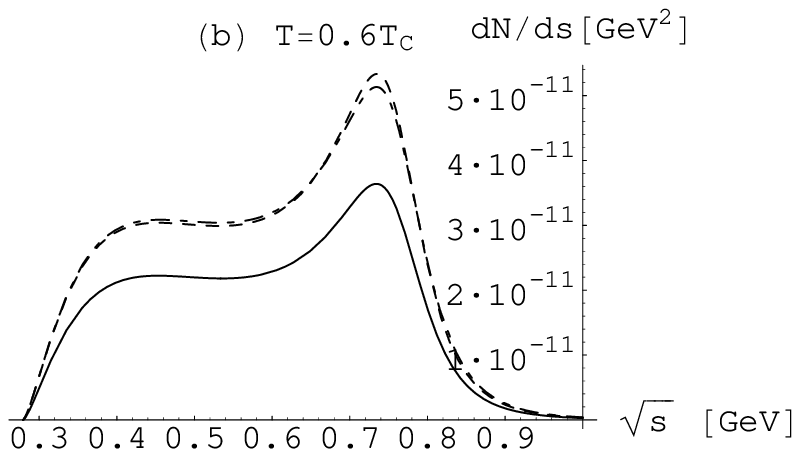}
\\
\includegraphics[width = 5cm]{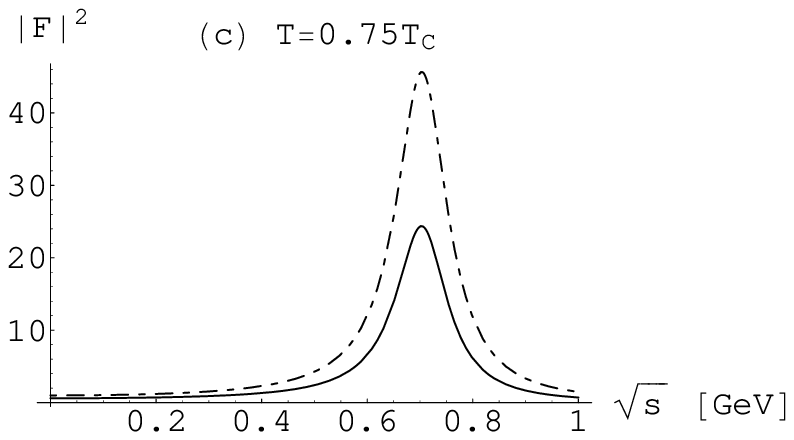}
\hspace*{0.5cm}
\includegraphics[width = 5cm]{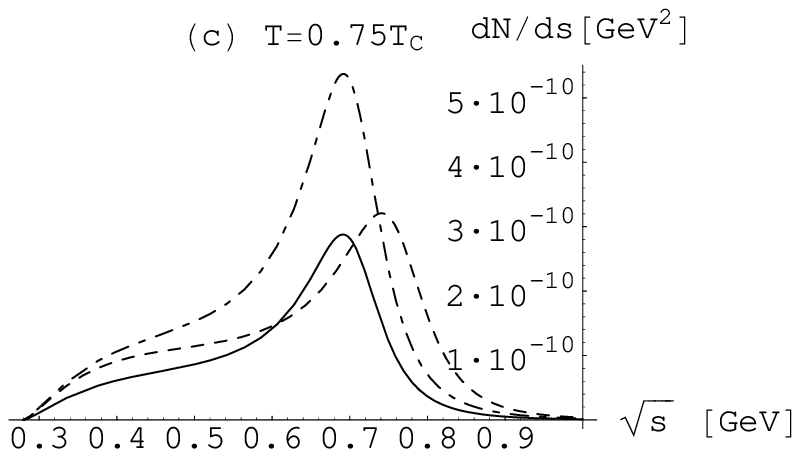}
\\
\includegraphics[width = 5cm]{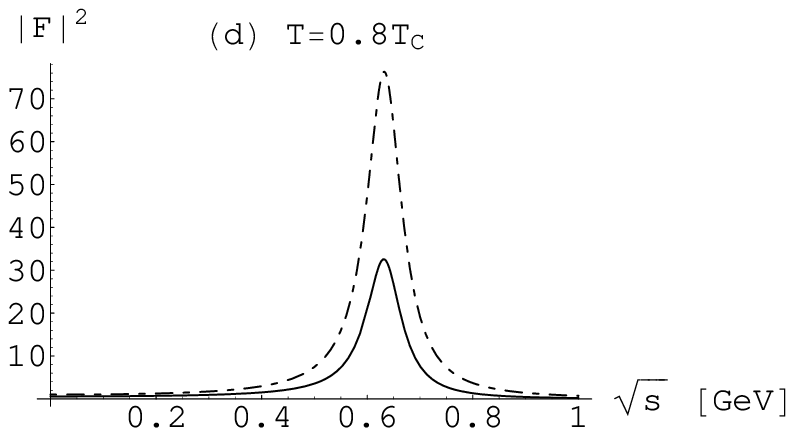}
\hspace*{0.5cm}
\includegraphics[width = 5cm]{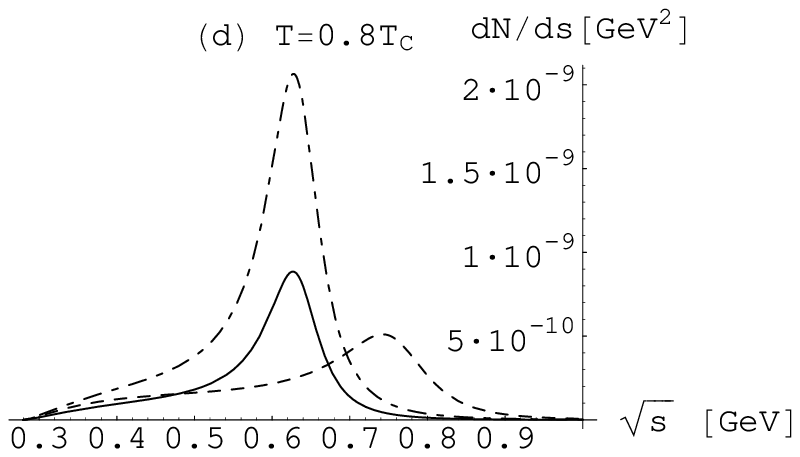}
\\
\includegraphics[width = 5cm]{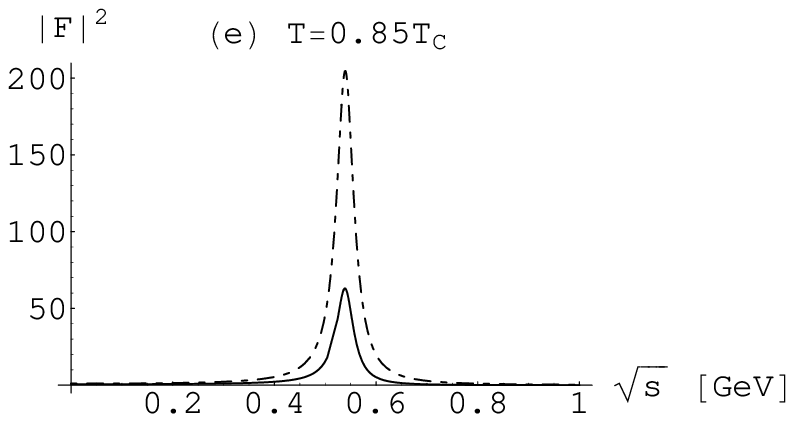}
\hspace*{0.5cm}
\includegraphics[width = 5cm]{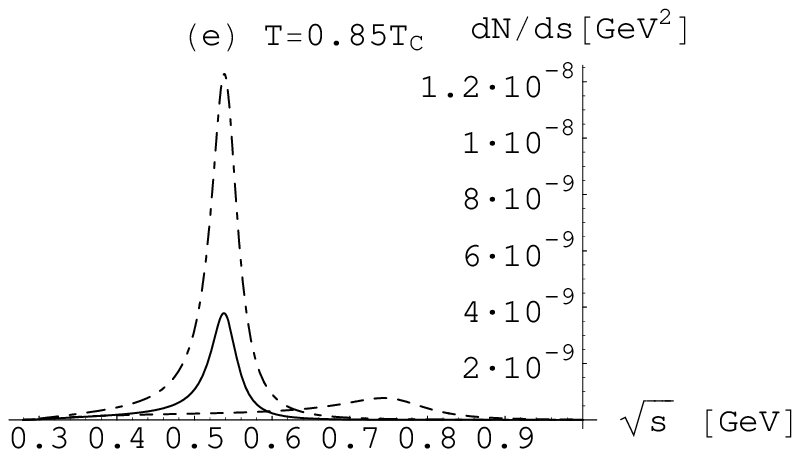}
\end{center}
\caption{
Electromagnetic form factor of the pion (left) and
dilepton production rate (right) as a function of the invariant 
mass $\sqrt{s}$ for various temperatures.
The solid lines include the effects of the violation of the VD.
The dashed-dotted lines correspond to the analysis assuming
the VD. In the dashed curves in the right-hand figures, 
the parameters at zero temperature were used.
}
\label{fig:dl}
\end{figure}

As noted,
the VM leads to the strong violation of the vector dominance (VD) 
(indicated by ``$\Slash{\rm VD}$'') 
near the chiral symmetry restoration point, which can be traced 
through the Wilsonian matching and the renormalization group evolutions.
Figure~\ref{fig:dl} shows the form factor and the dilepton 
production rate integrated over three-momentum,
in which the results with VD and $\Slash{\rm VD}$
were compared.
It can be easily seen that the $\Slash{\rm VD}$ gives a reduction
compared to the case with keeping the VD.
Below $T_f$ there are no much differences, while above $T_f$
a shift of the $\rho$ meson mass to lower-mass region can be seen
since the intrinsic temperature effects are turned on.
The form factor, which becomes narrower with increasing
temperature due to the dropping $m_\rho$, exhibits an obvious
discrepancy between the cases with VD and $\Slash{\rm VD}$.
The production rate based on the VM 
(i.e., the case with $\Slash{\rm VD}$) is suppressed compared
to that with the VD.
One observes that the suppression is more transparent 
for larger temperature.

As one can see in (c), the peak value of the rate
predicted by the VM
in the temperature region slightly above the flash temperature
is even smaller than the one obtained by the vacuum parameters,
and the shapes of them are quite similar to each other.
This indicates that it might be difficult to measure the 
signal of the dropping $\rho$ experimentally, if this
temperature region is dominant in the evolution of the fireball.
In the case shown in (d), on the other hand,
the rate by VM 
is enhanced by a factor of 
about two compared with the one by the vacuum $\rho$.
The enhancement becomes prominent near the critical temperature
as seen in (e).
These imply that we may have a chance to discriminate the
dropping $\rho$ from the vacuum $\rho$.


\setcounter{equation}{0}
\section{Summary and discussions}
\label{sec:sum}

We explored the electromagnetic form factor of pion and
the dilepton production rate in the vector manifestation (VM)
of chiral symmetry using the hidden local symmetry (HLS) theory
at finite temperature.
This framework involves the intrinsic temperature effect
of the Lagrangian parameters which leads to the dropping $m_\rho$
as the VM.
The form factor and the dilepton production rate receive 
non-negligible contributions from
the intrinsic effects.

A {\it naive} dropping $m_\rho$ formula, i.e., $T_f = 0$, 
as well as VD in hot/dense matter are sometimes used for
theoretical implications of the data.
As we have shown here, the intrinsic temperature effects
together with the violation of the VD give a clear difference
from the results without including those effects.
It may be then expected that a field theoretical analysis
of the dropping $\rho$ 
and a reliable comparison to dilepton measurements will provide 
an evidence for the in-medium hadronic properties associated with 
the chiral symmetry restoration, 
if complicated hadronization processes do not wash out those changes.

Our analysis can be applied to a study at finite density.
Especially to study under the conditions for CERN/SPS and
future GSI/FAIR would be an important issue.
In such a dense environment, the particle-hole configurations
with same quantum numbers with pions and $\rho$ mesons are 
crucial~\cite{FP}.
The violation of the VD has been also presented at finite density
in the HLS theory~\cite{HKR:VM}.
Therefore the dilepton rate as well as the form factor will be
much affected by the intrinsic density effects and be reduced
above the ``flash density''.
Nuclear many-body effects may provide a broadening and 
it would be important to study what happens on the dilepton
production if one includes both mass shift and collisional
broadening.

Recently the chiral perturbation theory with including vector
and axial-vector mesons as well as pions has been constructed%
~\cite{HS:GHLS} based on the generalized HLS.
In this theory the dropping $\rho$ and $A_1$ meson masses
were formulated and it was shown that the dropping masses
are related to the fixed points of the RGEs, one of which gives
a VM-type restoration and that the VD is strongly violated
also in this case.
It was proposed~\cite{BLR} that the dropping axial-vector mesons
can explain the anomalous $\rho^0/\pi^-$ ratio measured in 
peripheral collisions by STAR~\cite{STAR}.
Based on a field theoretical way,
inclusion of the effect of $A_1$ meson will be interesting.

\section*{Acknowledgments}

The work of C.S. was supported in part by the Virtual Institute
of the Helmholtz Association under the grant No. VH-VI-041.
The work of M.H. 
is supported in part by the Daiko Foundation \#9099, 
the 21st Century
COE Program of Nagoya University provided by Japan Society for the
Promotion of Science (15COEG01), and the JSPS Grant-in-Aid for
Scientific Research (c) (2) 16540241.


\end{document}